\documentclass[aps,prl,twocolumn,notitlepage,showpacs,floatfix,superscriptaddress,10pt, longbibliography]{revtex4-2}
\usepackage{listings}
\usepackage{dcolumn}
\usepackage{tabularx}
\usepackage{bm}
\usepackage{soul}
\usepackage{svg}
\usepackage{amsmath,amssymb,graphicx}
\usepackage{lipsum}
\usepackage[colorlinks=true,citecolor=blue,urlcolor=blue,linkcolor=blue]{hyperref}
\usepackage{environ}
\usepackage[normalem]{ulem}
\usepackage{tcolorbox}
\usepackage{orcidlink}
\usepackage{chemformula}
\usepackage{float}
\tcbuselibrary{listings,breakable,xparse,skins}

\makeatletter
\DeclareUnicodeCharacter{0301}{\'}

\begin{document}
\title{Strain effects on the fluctuation properties in noncollinear antiferromagnets: a first-principles and macrospin-based study }
\author{Mohammad M. Rahman\orcidlink{0000-0001-6108-6379}}
\affiliation{Associate, Physical Measurement Laboratory, National Institute of Standards and Technology, Gaithersburg, Maryland, USA}
\affiliation{Department of Chemistry and Biochemistry, University of Maryland, College Park, Maryland, USA }
\author{Farzad Mahfouzi\orcidlink{0000-0001-9297-3886}}
\affiliation{Associate, Physical Measurement Laboratory, National Institute of Standards and Technology, Gaithersburg, Maryland, USA}
\affiliation{Department of Chemistry and Biochemistry, University of Maryland, College Park, Maryland, USA }
\author{Matthew W. Daniels\orcidlink{0000-0002-3390-4714}}
\affiliation{Physical Measurement Laboratory, National Institute of Standards and Technology, Gaithersburg, Maryland, USA }
\author{Mark D. Stiles\orcidlink{0000-0001-8238-4156}}
\affiliation{Physical Measurement Laboratory, National Institute of Standards and Technology, Gaithersburg, Maryland, USA }
\date{\today}
\begin{abstract}
We present a theoretical investigation of epitaxial strain effects on the magnetic fluctuation properties of Mn$_3$Sn noncollinear antiferromagnets. Employing density functional theory (DFT), we uncover significant strain-induced modifications to key magnetic parameters, including magnetic anisotropy and both bilinear and biquadratic exchange interactions.  Our findings reveal that the biquadratic exchange, often neglected, plays a crucial role in defining the magnetic energy landscape and its response to strain.  These microscopic changes directly impact the energy barriers governing magnetic switching, thereby influencing thermal stability and fluctuation rates. Using macrospin-based simulations based on DFT-derived parameters, we provide a quantitative analysis of the macroscopic magnetic fluctuations influenced by these microscopic interactions. These insights are particularly relevant for applications requiring precisely controlled magnetic behavior, such as hardware for probabilistic computing.
\end{abstract}

\maketitle
\section{Introduction}
Magnetic materials are characterized by long-range correlated ground states, where constituent magnetic moments align collectively. These alignments often occur along preferred directions dictated by magnetic anisotropy---an effective energy landscape arising fundamentally from relativistic spin-orbit coupling interacting with the local crystal field environment, thus linking spin orientation to the lattice structure \cite{blundell2001magnetism, cullity2011introduction}. At finite temperatures ($T$), a dynamic competition ensues: the magnetic anisotropy energy barrier ($E_B$), which favors alignment along these easy axes, contends with thermal energy ($k T$), which promotes disorder through random fluctuations \cite{brown1963thermal, coffey2012thermal}, where $k$ is the Boltzmann constant. The ratio $\Delta = E_B / (k T)$, commonly known as the thermal stability factor, quantifies the resilience of the ordered state against such thermal agitation \cite{brown1963thermal, coffey2012thermal, weller2000high}. This parameter critically governs the characteristic timescale over which a magnetic configuration persists before potentially transitioning, often following relations like the N\'eel-Arrhenius law \cite{neel1949theorie, brown1963thermal}. These fluctuation dynamics determine a material's suitability for diverse technological applications, ranging from stable magnetic memory devices requiring large $\Delta$ for long-term data retention \cite{chappert2007emergence, weller2000high} to potentially fast computing elements leveraging controlled rapid fluctuations \cite{borders2019integer, sutton2017intrinsic, camsari2017stochastic}. While the behavior and dynamics of conventional ferromagnets and collinear antiferromagnets are relatively well-understood, technological progress increasingly demands novel material platforms exhibiting unique magnetic behaviors, necessitating a thorough understanding of their distinct properties and underlying physics.

Antiferromagnetic (AFM) materials hosting cluster magnetic multipoles, particularly the octupolar order found in noncollinear systems like Mn$_3$X (X = Sn, Ge, Ga, Pt, Ir, etc.), are emerging as an intriguing material class for novel spintronic applications. These materials feature noncollinear spin configurations, such as the 120$^\circ$ or 240$^\circ$ arrangements driven by magnetic frustration \cite{nakatsuji2015large, nayak2016large, kiyohara2016giant, surgers2014large, liu2017anomalous, shukla2025spintronic}. While possessing negligible net magnetization, their unique crystal and magnetic symmetries enable several macroscopic phenomena that are typically absent in conventional antiferromagnets. Crucially, the collective magnetic state can be effectively described by a ferroic ordering of a cluster magnetic octupole moment, which acts as the order parameter \cite{nakatsuji2015large, suzuki2017cluster, liu2017anomalous}. This octupole transforms under similar symmetry operations as an ordinary ferromagnetic dipole moment \cite{suzuki2017cluster}. This symmetry enables a large anomalous Hall effect, originating from the large Berry curvature in the electronic bands \cite{nakatsuji2015large, nayak2016large, chen2021anomalous}, and finite tunneling magnetoresistance in corresponding tunnel junctions, originating from momentum-dependent spin polarization allowed by the space group \cite{dong2022tunneling, chen2023octupole,Shi2024}. Consequently, these octupolar antiferromagnets offer a unique combination: the terahertz dynamics and robustness against external field perturbations similar to conventional AFMs, merged with efficient electrical detection and readout via the anomalous Hall effect and tunneling magnetoresistance, similar to ferromagnets but typically not observed in conventional collinear AFMs. This positions octupolar AFMs as holding significant promise for advancing AFM-based applications in efficient memory, computing, and other magnetic device technologies.

Despite the significant promise of octupolar antiferromagnets outlined above, realizing their full potential faces challenges, particularly concerning the control and understanding of their fluctuation dynamics, where epitaxial strain emerges as a critical, often unavoidable factor \cite{sato2023thermal, higo2018anomalous, deng2022effect, he2024magnetic, shukla2024impact, theuss2022strong, torres2022anomalous}. For instance, in the hexagonal D0$_{19}$ structure common to Mn$_3$X compounds, the Mn atoms form Kagome lattices. Intrinsic magnetic frustration within these ideal Kagome layers, combined with the $C_{3v}$ symmetry, theoretically stabilizes six equivalent states corresponding to different orientations of the octupole moment within the Kagome plane separated by ultra-low energy barriers \cite{liu2017anomalous, pal2022setting}. The combination of inherent low-barrier along with exchange-driven dynamics intrinsic to antiferromagnets suggests that fast switching dynamics may be readily accessible. However, experimental measurements on thin-film samples of these materials often reveal significantly higher effective energy barriers and thermal stability factors than predicted by this idealized picture \cite{sato2023thermal}. This discrepancy is widely attributed to epitaxial strain \cite{sato2023thermal, takeuchi2021chiral, konakanchi2025electrically, he2024magnetic}, inevitably present in thin films due to lattice mismatch with the underlying substrate during growth. Such strain breaks the inherent structural symmetries of the Kagome lattice (e.g., reducing three-fold rotational symmetry), which in turn modifies the magnetic interactions---including exchange interactions (both bilinear and potentially higher-order like biquadratic) and magnetocrystalline anisotropy. This reshaping of the magnetic energy landscape lifts the ground state degeneracy and alters the energy barriers between states, consequently modulating the fluctuation timescales and overall thermal stability. Therefore, achieving precise control over the dynamics of these octupolar antiferromagnets necessitates a quantitative, microscopic understanding of how strain modulates these fundamental magnetic interactions. So far, investigations into strain effects have often relied on indirect methods, such as fitting experimental trends to phenomenological models with effective parameters \cite{liu2017anomalous, takeuchi2021chiral, shukla2024impact, yamane2019dynamics}. 

In this work, we make the following contributions: we use first-principles density functional theory (DFT) to extract the behavior of classical magnetic Hamiltonian parameters under substrate-induced strain, specifically targeting configurations where the Kagome plane is oriented perpendicular to the growth direction. This orientation enables spin-torque control of the magnetic order and has been adopted in recent experiments \cite{takeuchi2021chiral, tsai2020electrical, yoon2023handedness}. Other configurations are also expected to exhibit similar strain-induced effects. We find that, in addition to the isotropic Heisenberg exchange, the biquadratic exchange plays a key role in determining both the magnetic ground state and the excitation dynamics. In particular, we demonstrate that the strain-induced symmetry reduction generates a competition between the Heisenberg and biquadratic exchange terms, which collectively govern the thermal stability of these magnets. Employing a simple compact model based on this Hamiltonian, we compute the energy barriers and thermal stability of devices composed of Mn$_3$Sn. The calculated thermal stability for the expected lattice mismatch with the substrate is consistent with that inferred from experiments. Finally, we use the theoretical framework to predict the fluctuation spectra of the octupole moment.

\begin{figure}[ht]
\includegraphics[width=0.49\textwidth]{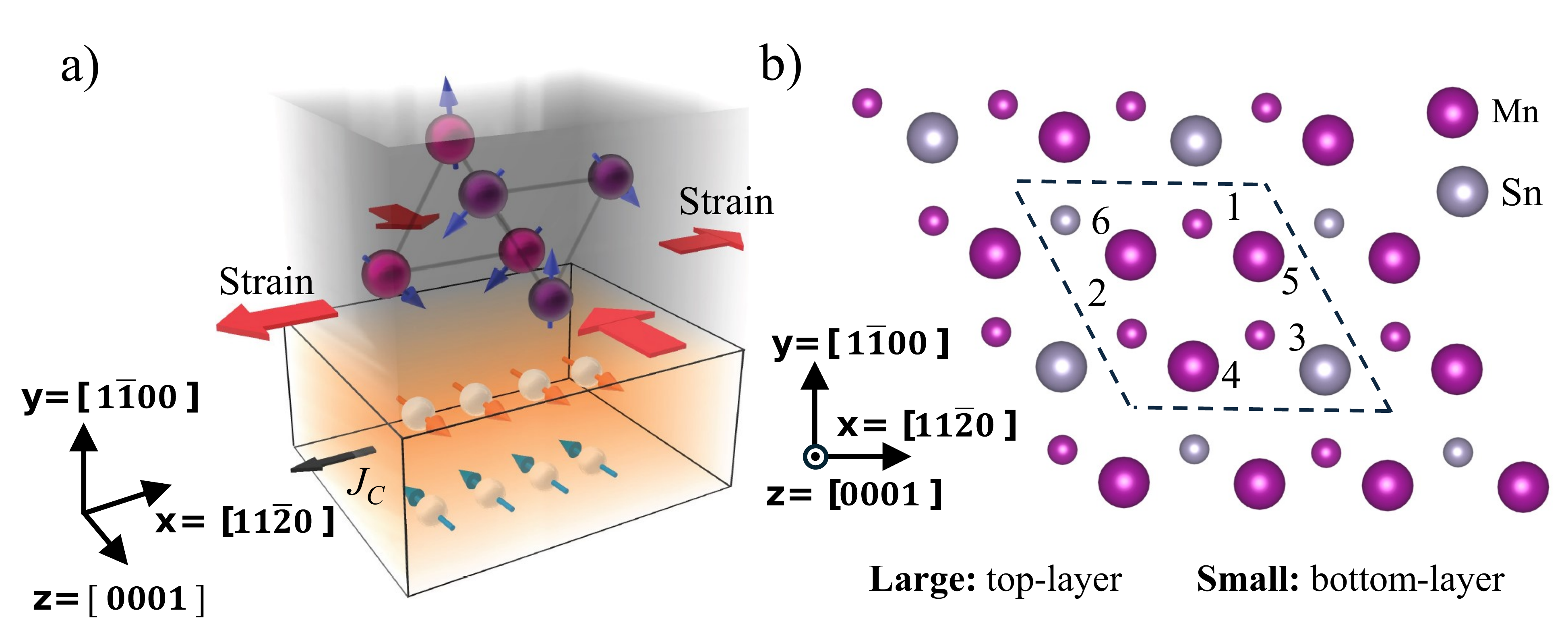}
\caption{(a) Schematic orientation of a Mn$_3$Sn film structure considered in this study. The crystallographic $c$-axis (normal to the Kagome planes) is oriented in the substrate surface plane. This orientation allows experimental spin-torque control of the magnetic order: an in-plane current ($J_C$) in an adjacent strong spin-orbit coupling material (lower layer in schematic) generates a transverse spin current via the spin-Hall effect. The injection of the spin current into Mn$_3$Sn results in spin-orbit torque that reorients the octupolar order, highlighting the utility of this perpendicular Kagome plane geometry (explored here as a proof-of-concept, with other orientations also expected to show similar strain-induced effects).  Epitaxial strain, simulating lattice mismatch, is applied along the in-plane $[11\bar{2}0]$ direction; to preserve approximate volume, a corresponding compressive/expansive strain is simultaneously applied along the out-of-plane $c$-axis, as schematically indicated in panel (a).
(b)  Crystal structure of Mn$_3$Sn. The D0$_{19}$ (P6$_3$/mmc) unit cell consists of ABAB... stacking with a weakly distorted hexagonal close-packed layers. Mn atoms (pink spheres at $z=1/4$, garnet spheres at $z=3/4$ fractional heights along the $c$-axis) form the characteristic Kagome planes; Sn atoms (smaller grey spheres) are also positioned within these $z=1/4$ and $z=3/4$ layers. Unstrained lattice parameters are $a = b = 0.567$~\text{nm} and $c = 0.453$~\text{nm}. Epitaxial strain, simulating lattice mismatch, is applied along the in-plane $[11\bar{2}0]$ direction. This strain breaks the inherent $C_{3v}$ symmetry of the ideal Kagome lattice, leading to anisotropic magnetic interactions and influencing the triangular spin configuration observed in the $xy$-plane.}
\label{fig:structure}
\end{figure}

\section{Magnetic structure and effective free energy of {\texorpdfstring{\(\text{Mn}_3\text{Sn}\)}{Mn3Sn}} }
The magnetic behavior of Mn\textsubscript{3}Sn, a hexagonal antiferromagnet is governed by a complex interplay of interactions among the Mn moments arranged on a Kagome-like lattice. The interaction between the spins can be described by
\begin{align}
H &= \sum_{i > j} J_{ij}\, \bm{S}_i\cdot \bm{S}_j + \sum_{i> j} B_{ij}\ \left(\bm{S}_i\cdot \bm{S}_j\right)^2  \notag\\
&+ \sum_{i> j} D_{ij} \hat{z} \cdot (\bm{S}_i\times \bm{S}_j) - K_1 \sum_i (\bm{S}_i\cdot \bm{u}_i)^2,
\label{eq:F0Hamiltonian} 
\end{align}
where, $J_{ij}$ is the bilinear Heisenberg exchange, $B_{ij}$ is the biquadratic exchange, ${D}_{ij}$ is the Dzyaloshinskii-Moriya interaction (DMI), $K_1$ is the site-dependent uniaxial magnetocrystalline anisotropy with local easy axes $\bm{u}_i$. The directions of the anisotropy are dictated by the local crystal symmetry and spin-orbit coupling effects primarily originating from the Sn atoms. The applied strain in this study is applied along $\hat{x}$ with a corresponding change along $\hat{z}$ (in the surface plane) to keep the unit cell volume constant. Here, we focus on the collective excitations ($q=0$) of the magnetic order, the equivalent of uniform modes in ferromagnets. For uniform excitations, all of the pairwise exchange interactions between different unit cells ($J_{ij}$ and $B_{ij}$) effectively sum to determine net exchange fields within the unit cell. Hereafter, we restrict ourselves to these net interactions, which dictate the stability and low-energy dynamics of the uniform magnetic configuration. 

The inherently weak spin–orbit coupling in Mn\textsubscript{3}Sn gives rise to a clear hierarchy among the effective fields experienced by the spins, with the exchange interactions being the strongest, followed by DMI, and anisotropy being the weakest \cite{liu2017anomalous}. The Heisenberg exchange $J_{ij}$, which is the strongest, is primarily responsible for establishing the characteristic 120$^\circ$ or 240$^\circ$ noncollinear triangular spin configurations in the Kagome plane~\cite{nakatsuji2015large, liu2017anomalous}. Although the biquadratic exchange term is often ignored in analyses of Mn$_3$Sn spin dynamics, we demonstrate that its inclusion is essential to accurately capture the spin behavior and the associated energy landscape, and is fairly common for Mn-based compounds \cite{harris1963biquadratic, rodbell1963biquadratic, gaulin1986biquadratic, amirabbasi2024effect, fedorova2015biquadratic, trukhanova2024new}. 
\begin{figure}[htbp]
  \centering
  \includegraphics[width=0.49\textwidth]{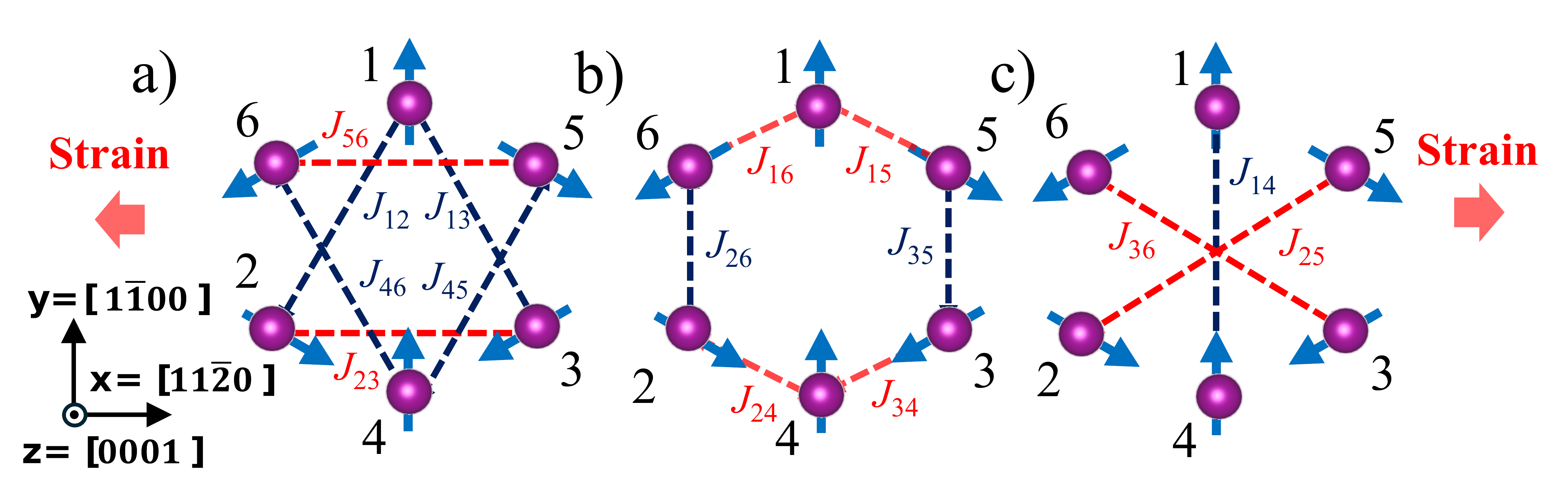}
\caption{Illustration of the distinct pairwise bilinear ($J_{ij}$) exchange interactions within the Mn$_3$Sn unit cell under epitaxial strain. Strain along the $[11\bar{2}0]$ direction breaks the three-fold rotational symmetry, making previously equivalent bonds (illustrated using red and dark blue dashed lines in a, b and c) distinct. Similar modifications in biquadratic exchange ($B_{ij}$) are also considered in the study.}
\label{fig:introducing_nu}
\end{figure}

For the experimentally relevant regime, where bilinear exchange is greater than biquadratic exchange, the system robustly maintains its triangular spin ordering. The opposite scenario, however, could lead to out-of-plane ``umbrella" distortions, in which all spins tilt out of plane, not typically observed as the ground state in Mn\textsubscript{3}Sn. The DMI term ${D}_{ij}$ lifts the chiral degeneracy between the clockwise (120$^\circ$) and counter-clockwise (240$^\circ$) triangular spin arrangements, energetically selecting the latter in Mn\textsubscript{3}Sn. Finally, the single-ion anisotropy $K_1$, acting with the exchange interaction, breaks the continuous $U(1)$ rotational symmetry of the in-plane spin moments. This symmetry breaking leads to six discrete, energetically equivalent ground states in the unstrained crystal. The energy barrier separating these states is expected to scale roughly as $\sim K_1^3/J^2$~\cite{liu2017anomalous}, suggesting a potentially shallow energy landscape conducive to efficient manipulation of the magnetic order. 

A central aim of this paper is the first-principles determination of these Hamiltonian parameters ($J_{ij}$, $B_{ij}$, and $K_1$) using density functional theory (DFT). While the Dzyaloshinskii–Moriya interaction plays a role in selecting the spin chirality, its influence on the strain-dependent static and dynamic properties, beyond setting the chirality, is negligible; hence, we did not include DMI in our calculations—a choice justified \textit{a posteriori} and validated in subsequent analysis. 

The pristine (unstrained) structure exhibits $C_{3v}$ symmetry within the Kagome layers, implying that several sets of atom pairs are equivalent. These include specific in-plane pairs (e.g., $(1,2) \Leftrightarrow (1,3) \Leftrightarrow (2,3)$ within one layer, and similarly for $(4,5) \Leftrightarrow (4,6) \Leftrightarrow (5,6)$ in the other), cross-plane pairs (e.g., $(1,5) \Leftrightarrow (1,6) \Leftrightarrow \dots$), and diagonal pairs $(1,4) \Leftrightarrow (2,5) \Leftrightarrow (3,6)$ (see Fig.~\ref{fig:structure} for atom numbering). However, strain along the $[11\bar{2}0]$ direction breaks this three-fold rotational symmetry and breaks the symmetry between the lateral and vertical bonds (see Fig.~\ref{fig:introducing_nu}). Consequently, both the bilinear ($J_{ij}$) and, significantly, the biquadratic ($B_{ij}$) exchange parameters become different for differently oriented pairs. We therefore quantify the bond-selective exchange mismatch by taking explicit differences between exchange parameters of bonds that are symmetry-equivalent in the unstrained lattice (e.g. $J_{12}$ and $J_{23}$) but are split by strain (e.g. $\Delta J \equiv J_{12} - J_{23}$).

\begin{figure}[htbp]
    \centering
    \includegraphics[width=0.998\linewidth]{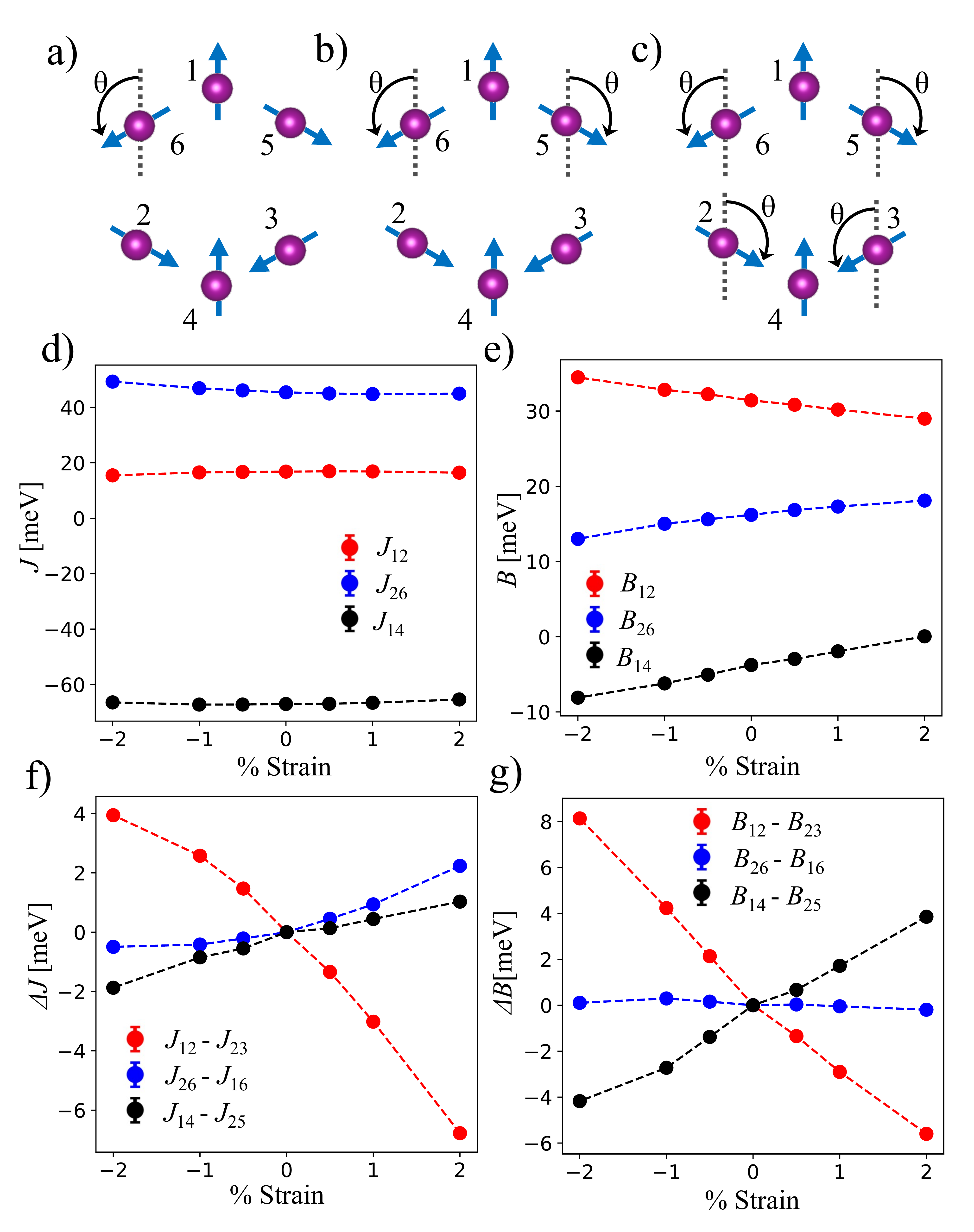}
\caption{Extraction of pairwise magnetic interactions from DFT calculations. (a--c) Schematics of typical spin configurations used for energy mapping: (a) single-spin, (b) two-spin synchronous, and (c) four-spin synchronous rotations. These configurations generate the energy dataset for fitting the Hamiltonian (Eq.~\eqref{eq:F0Hamiltonian}). A total of 14 independent configurations with spins rotated in 20$^\circ$ steps for a total of 252 separate self-consistent field calculations were performed for each strain value to generate the dataset. (d,e) DFT-extracted bilinear ($J_{ij}$) and biquadratic ($B_{ij}$) exchange constants as a function of epitaxial strain. Indices $i,j$ refer to atom numbering. (f,g) Illustration of strain-induced symmetry breaking in selected exchange parameters. $\Delta J_{ij}$ and $\Delta B_{ij}$ denote the strain-induced changes in the bilinear and biquadratic exchange parameters, respectively. For example, in the unstrained lattice all nearest‐neighbor exchange constants are equal to each other, $J_{12} = J_{13} = J_{23}$. Under epitaxial strain, the three-fold symmetry is lifted and one group of bonds shift by an amount $\Delta J = J_{12} - J_{23}$ while $J_{12} = J_{13}$ (see Fig.~\ref{fig:introducing_nu} for the symmetry equivalent bonds in presence of strain). Error bars indicate the one standard deviation fitting uncertainties and are not visible where they are smaller than the marker size. }
\label{fig:DFT_mapping_schematic}
\end{figure}

Equipped with the above understanding, we now proceed to first compute the exchange interactions. For that, we perform spin-constrained density functional theory (DFT) calculations using the Quantum ESPRESSO package \cite{giannozzi2017advanced, giannozzi2009quantum} (computational details are provided in Appendix~I). The pairwise exchange interactions, both bilinear ($J_{ij}$) and biquadratic ($B_{ij}$), are systematically extracted by fitting to the total energies of various noncollinear spin configurations. These configurations are generated by selectively rotating the magnetic moments of one, two, or four Mn spins relative to their neighbors while keeping other moments fixed, as illustrated in Fig.~\ref{fig:DFT_mapping_schematic}(a-c). 

The fitted parameters are shown in Fig.~\ref{fig:DFT_mapping_schematic}(d–g). As expected, the exchange parameters for the unstrained system exhibit symmetry, as indicated by the negligible values of $\Delta J$ and $\Delta B$ near 0~\% strain. We find that the bilinear Heisenberg exchange alone is insufficient to capture the relevant physics underlying the energy evolution (see Appendix I). However, incorporating two-sublattice biquadratic interactions ($B_{ij}$) is sufficient to accurately reproduce the energy landscape. Additional higher-order terms (e.g., three-sublattice fourth-order interactions) were tested but they did not significantly improve the quality of the fit. Consequently, we restrict our effective Hamiltonian [Eq.~\eqref{eq:F0Hamiltonian}] to include only pairwise bilinear and biquadratic interactions. 
We highlight that the values shown correspond to a single pair interaction, which includes contributions from all other crystallographically equivalent bonds in the lattice. Finally, we also note that our extracted parameters are consistent with those reported in Ref.~\cite{dos2021proper}, which analyzed an effective three-atom sublattice model for the unstrained system.

\begin{figure}[ht]
\includegraphics[width=0.48\textwidth]{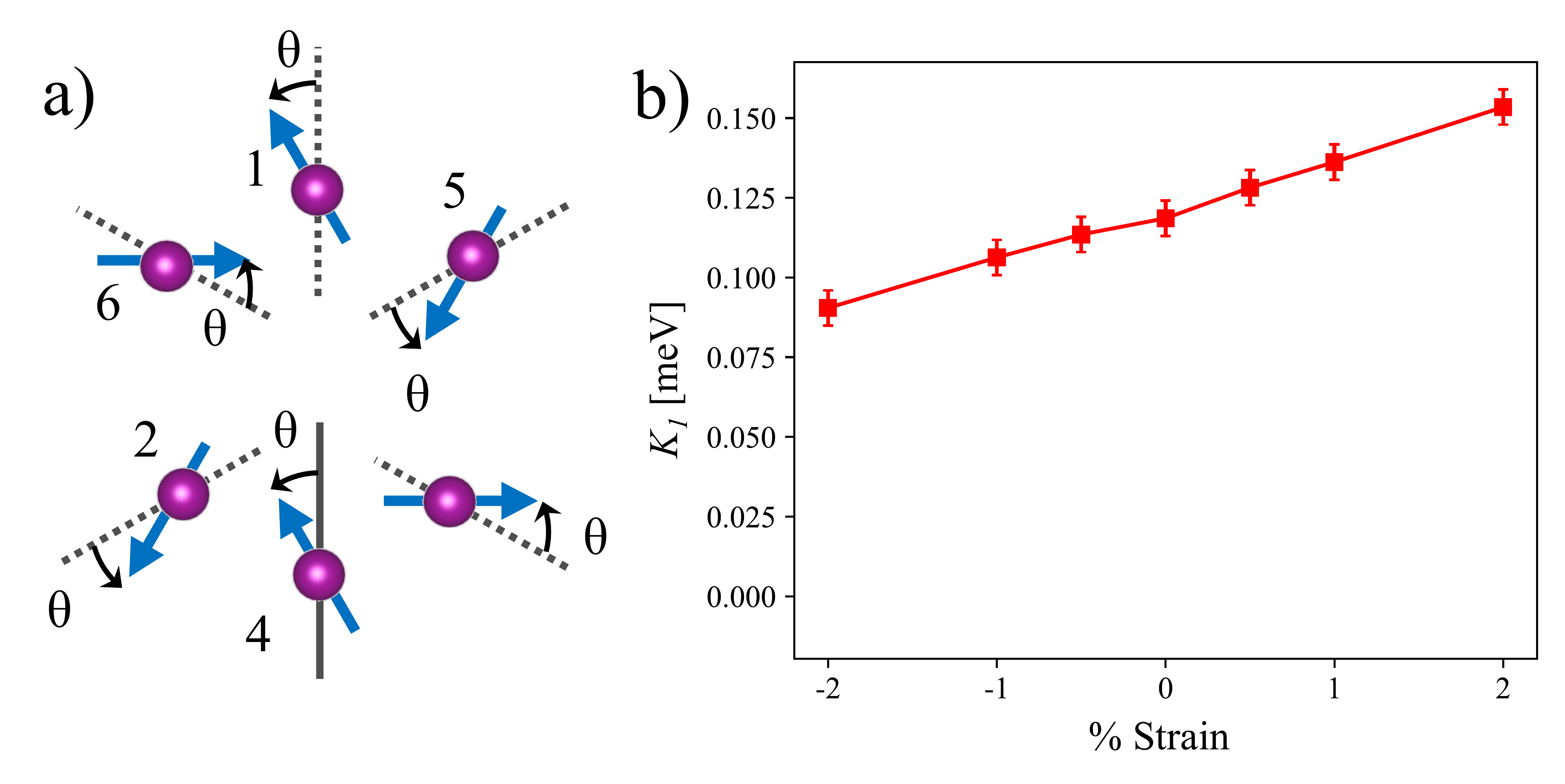} 
\caption{DFT calculation of magnetocrystalline anisotropy in Mn$_3$Sn. (a) Schematic showing the rigid rotation by an angle $\theta = 20^\circ$ of the entire $120^\circ$ triangular spin structure relative to the crystal lattice, used to extract anisotropy energies from DFT calculations including spin-orbit coupling. A total of 18 SCF calculations per strain value were performed. Note that the spin configurations used here have the opposite chirality of those close to the ground state considered in Fig.~\ref{fig:introducing_nu} and Fig.~\ref{fig:DFT_mapping_schematic}. (b) Extracted single-ion anisotropy constants ($K_1$) as a function of applied epitaxial strain. 
Error bars represent one standard deviation uncertainties extracted from the fitting procedure.}
\label{fig:anis_dft}
\end{figure}

Next, we turn to the site-dependent magnetocrystalline anisotropy energy, focusing on the leading uniaxial anisotropy constant $K_1$. The corresponding local magnetic easy/hard axes, $\bm{u}_i$, for the $i$th Mn moments are dictated by the crystal symmetry and the local environment created by the surrounding Sn atoms, aligning along directions such as $[21\bar{1}0]$ and its symmetry-equivalent orientations within the Kagome plane for the different Mn sublattices. To compute $K_1$, we employ spin-constrained DFT calculations with spin-orbit coupling activated. The total energy is calculated as the entire $120^\circ$ triangular spin structure (representing the established noncollinear order, see Fig.~\ref{fig:anis_dft}(a)) is rigidly rotated with respect to the crystal lattice. Here, maintaining the opposite chirality is crucial as it effectively isolates the magnetocrystalline anisotropy contribution: the energy change due to isotropic exchange interactions ($J_{ij}$ and $B_{ij}$) remains null, and the contribution from the Dzyaloshinskii-Moriya interaction stays constant so does not affect the results. The variation in total energy as a function of the structure's orientation is fitted to the anisotropy terms, as shown in Fig.~\ref{fig:anis_dft}(b).
Strain can reorient the individual anisotropy axes $\bm{u}_i$, which in earlier model calculations\cite{he2024magnetic} was treated by including an anisotropy term of the form $K_2 (\bm{S}_i\cdot \bm{x})^2$, where $\bm{x}$ is the strain direction. We found that for the moderate strains we consider, the coefficient $K_2$ is statistically indistinguishable from zero and does not play a significant role in the dynamics we consider below.

\section{Low energy equations of motion of the octupole}
With the microscopic Hamiltonian parameters established, we now focus on deriving the equations governing the low-energy dynamics of the collective cluster octupole moment. The dynamics of the six Mn sublattices can be effectively reduced to an equivalent three-spin model for these low-energy modes. This reduction is justified by the presence of strong effective ferromagnetic coupling that pairs specific Mn spins across the two distinct Kagome planes (e.g., atoms 1 with 4, 2 with 5, and 3 with 6, as per the atomic indexing in Fig.~\ref{fig:structure}(b)), forming three effective macrospins whose slower, collective motion dictates the system's response (further details and justification are provided in Appendix~III, following a similar approach to Ref.~\cite{he2024magnetic}). Within the established hierarchy of interactions, and employing a perturbative treatment by adiabatically eliminating the fast optical modes, we arrive at a set of coupled differential equations for the two slow collective variables: $\phi_{\text{oct}}$, representing the uniform azimuthal orientation of the in-plane octupole component (e.g., measured from the $[11\bar{2}0]$ axis as depicted in Fig.~\ref{fig:structure}(b)), and $m_z$, representing the out-of-plane canting of the dipole moment of the combined Mn spins,
\begin{subequations}
\label{eq:final_eom_theta0_main}
\begin{align}
    \dot{m}_z + \alpha \dot{\phi}_{\text{oct}} &= \mu_0\gamma H_K^{\text{eff}}\sin(2\phi_{\text{oct}}),\\
   \text{and}\quad \dot{\phi}_{\text{oct}} - \alpha\dot{m}_z &= \mu_0\gamma H_{\perp}^\text{eff}m_z. 
\end{align}
\end{subequations}
Here $\gamma$ is the gyromagnetic ratio, $\alpha$ is the Gilbert damping parameter, and $\mu_0$ is the permeability of free space. $H_K^{\text{eff}}$ and $H_{\perp}^{\text{eff}}$ denote the effective anisotropy field and the effective demagnetizing fields, 
\begin{align}
 H_K^{\text{eff}} &= \frac{2 K(\Delta J - \Delta B)}{3 \mu_0 M_S (2B+\sqrt{3}\,D+J)}, \\
 \text{and}\quad H_{\perp}^{\text{eff}} &= \frac{ (3J - 3B + \sqrt{3} D)}{ \mu_0 M_S}, \label{eq:omega_perp_main}
\end{align}
where $M_S=6 \mu_B/V_{cell}$ is the saturation magnetization of the macrospin, and $V_{cell}$ denotes the unit cell volume. $J$ and $B$ represent the effective isotropic bilinear and biquadratic exchange parameters, respectively derived by considering the contributions from the main interactions (e.g. $J = J_{12} + J_{15} + J_{24} + J_{45}$ and similarly for $B$ using their symmetry-equivalent combinations). $\Delta J$ and $\Delta B$ quantify the relative changes in these exchange interactions due to the strain-induced symmetry breaking. $K=2K_1$ denotes the effective macrospin anisotropy parameter and $D$ represents the effective DMI constant connecting the macrospins, obtained from considering a homogeneous intralayer DMI coupling. Although DMI enters the expressions for $H_K^\text{eff}$ and $H_\perp^\text{eff}$, its magnitude is typically an order of magnitude smaller than $J$ or $B$ as reported in previous studies \cite{liu2017anomalous, shukla2024impact, konakanchi2025electrically, yamane2019dynamics} and confirmed by our own DFT mapping (results not shown). The dynamical system described by Eqs.~\eqref{eq:final_eom_theta0_main} is directly analogous to that of an easy-plane ferromagnet, where $\phi_{\text{oct}}$ and $m_z$ behave as conjugate dynamical variables. In this analogy, $H_K^{\text{eff}}$ acts as an effective in-plane anisotropy field, which is influenced by the interplay of uniaxial anisotropy $K_1$ and competition between the strain-induced exchange anisotropies. Meanwhile $H_{\perp}^\text{eff}$ determines the effective demagnetizing stiffness, resisting deviations of the Mn moments from the Kagome plane.

In the absence of strain, the symmetry of the lattice dictates that the anisotropic exchange terms $\Delta J$ and $\Delta B$ are zero. Consequently, the effective twofold anisotropy field $H_K^{\text{eff}}$ in Eq.~\eqref{eq:omega_perp_main} vanishes. In this symmetric limit, the dynamics are instead governed by the weaker, intrinsic six-fold magnetocrystalline anisotropy \cite{liu2017anomalous, yamane2019dynamics} that arises from the interplay of exchange and anisotropy. The energy barrier separating these states scales as $E_B =  K_1^3 V/12J^2$, which determines the thermal stability and relaxation dynamics at zero strain. 

\section{Fluctuation properties of the octupole}
\begin{figure*}[htbp]
  \centering
  \includegraphics[width=0.99\textwidth]{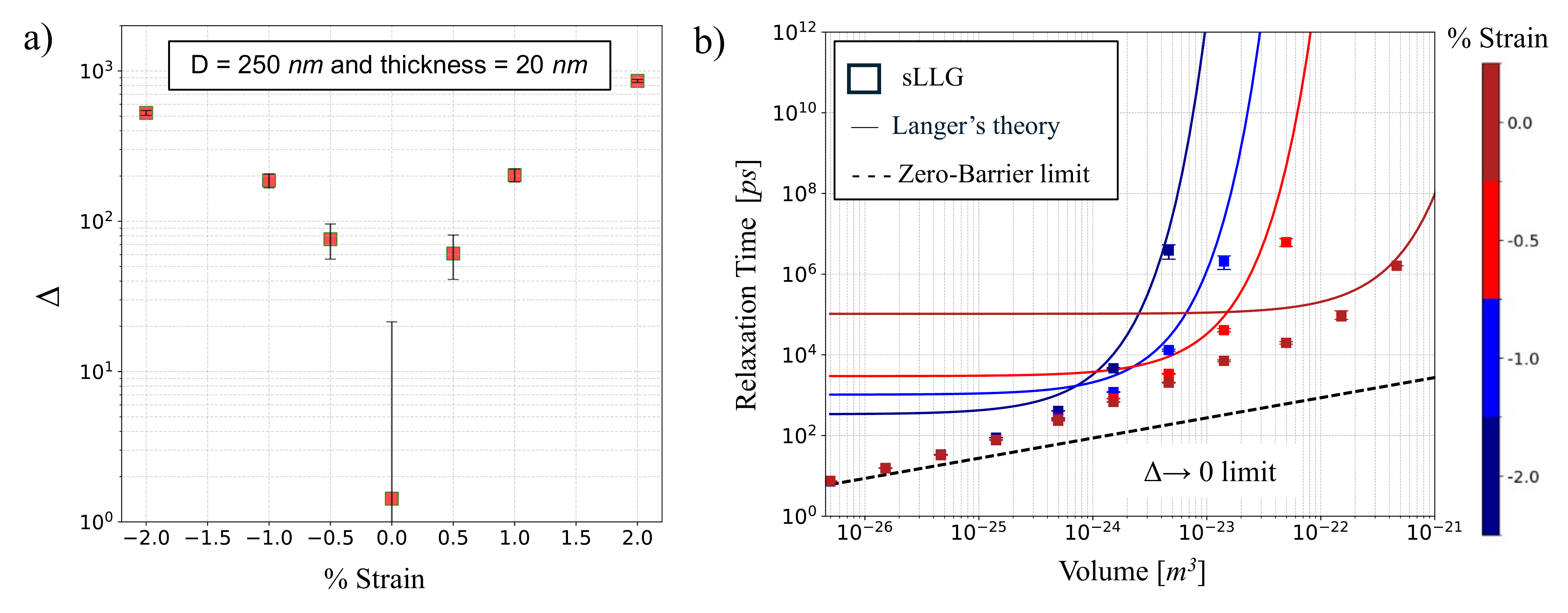}
    \caption{Strain and volume dependence of octupole fluctuation properties in Mn$_3$Sn. (a) Calculated thermal stability for octupole switching as a function of applied epitaxial strain, derived from the effective model with DFT parameters for the given volume of the magnet. For finite strain, the barrier is determined by the twofold anisotropy arising from symmetry breaking. At zero strain, this twofold term vanishes due to symmetry, and the thermal stability is instead governed by the underlying six-fold anisotropy of the unstrained lattice \cite{liu2017anomalous, yamane2019dynamics}. The error bars represent variations in $\Delta$ arising from the single standard deviation uncertainty in the DFT-derived magnetic parameters. (b) Octupole moment relaxation time ($\tau$) versus magnetic volume, comparing stochastic LLG simulations (symbols) with analytical models: Langer’s theory (solid line) for high barriers and a zero-barrier limit formula (dashed line) for low barriers. The relaxation time is defined as the $1/e$ decay time of the octupole autocorrelation function. Error bars on the sLLG data represent the single standard deviation uncertainty in the computed relaxation time.}
    \label{fig:barrier}
\end{figure*}

The easy-plane-like equations of motion in Eqs.~\eqref{eq:final_eom_theta0_main} and ~\eqref{eq:omega_perp_main} allows for a direct extraction of the thermal stability factor, $\Delta$. Physically, the minimum energy path connecting two adjacent energy minima corresponds to the uniform in-plane rotation of the collective octupole moment, $\phi_{\text{oct}}$. This theoretical framework allows for a quantitative comparison with experimental findings. For instance, a recent study by Sato et al.~\cite{sato2023thermal} investigated the thermal stability factor $\Delta$ in Mn$_3$Sn nanodots by measuring switching probabilities under pulsed magnetic fields. In particular, they computed $\Delta$ values and found signatures of two-fold anisotropy which can be interpreted as a direct manifestation of strain-induced symmetry breaking within the otherwise sixfold-symmetric Kagome lattice---a phenomenon that our combined DFT and micromagnetic model also predicts. 

As a first step towards making a connection with these experiments, we compute $\Delta$ using our model, which relies on the minimum energy path being a uniform in-plane rotation of the collective octupole moment--a pathway confirmed by our nudged elastic band calculations for a six-spin system using the Hamiltonian in Eq.~\eqref{eq:F0Hamiltonian}. The resulting trends, illustrated in Fig.~\ref{fig:barrier}(a), demonstrate that increasing compressive strain leads to a higher $\Delta$, consistent with the enhancement of the effective anisotropy barrier due to symmetry reduction. We highlight that the $\Delta$ values obtained from our model are comparable in magnitude to those reported by Sato et al.~\cite{sato2023thermal} for strain levels that are realistically expected in epitaxially grown Mn$_3$Sn/substrate stacks. This congruence provides a plausible microscopic explanation for the experimentally observed two-fold anisotropy and underscores the predictive power of our strain-dependent Hamiltonian.

We compute the rate of entropy generation when these magnets are connected to a thermal bath by numerically integrating the stochastic Landau-Lifshitz-Gilbert equation within an equivalent three-macrospin space (for details, see Appendix IV) to collect relevant statistics. These simulations provide a robust numerical benchmark for the octupole's fluctuation dynamics. 

The numerical behavior can be understood in two regimes analytically distinguished by the energy barrier $E_B$ seen by the octupole relative to thermal energy $k T$, as highlighted in recent work \cite{konakanchi2025electrically}.  In the low-barrier limit, the correlation decay can be described by analytical expressions derived for zero barrier dynamics in easy-plane magnets, capturing the rapid loss of correlation to $C(t) = \exp(-\gamma^2 H_\perp^{\text{eff}} k T t^2/ 2 M_S V)$. The dashed line in Fig.~\ref{fig:barrier}(b) represents this estimate and shows good agreement with simulations for small volumes. The loss of correlation stems from random thermal kicks deflecting the spin cluster out-of-plane; the effective perpendicular field $H_{\perp}^{\text{eff}}$ then induces random precessional dynamics of the in-plane octupole moment $\phi_{\text{oct}}$. As the magnetic volume $V$ increases, the impact of thermal fields (which scales as $1/V$) lessens. This leads to longer correlation times, as observed in the simulation data in Fig.~\ref{fig:barrier}(b).

On the other hand, when the energy barrier is significantly larger than $k T$ (characteristic of larger volumes or higher compressive strains), the system's long-time dynamics are dominated by thermally activated escape over this barrier. This regime is well-described by Langer's theory of reaction rates, which provides an expression for the rate constant of barrier crossing based on the properties of the energy landscape, such as the barrier height $E_B$, the instability (negative curvature) at the saddle point, and the curvatures at the energy minima. The average barrier crossing time under these conditions can be expressed as \cite{langer1969statistical, konakanchi2025electrically, desplat2020entropy} 
\begin{equation}
\tau = \frac{2\pi}{\lambda_+} \sqrt{\frac{\prod_j |\lambda_j^S|}{\prod_i \lambda_i^A}} \exp(\Delta)
\end{equation}
where the $\lambda_+$ denotes the unstable eigenmode at the saddle point and $\lambda_i^A$ and $\lambda_j^S$ represent the $i^\text{th}$ and $j^\text{th}$ curvature of the energy surface at the minima and saddle point, respectively. Our numerical results, presented in Fig.~\ref{fig:barrier}(b), corroborate this picture: as the volume increases, thereby increasing $E_B$, Langer's theory shows progressively better agreement with the simulation data. This transition is physically expected, as Langer's formulation accurately captures the Arrhenius-like exponential dependence of the escape rate on $\Delta$.

\section{Discussion}
While the current analysis of relaxation time is predominantly based on a macrospin model, in reality, magnetic reversal in larger films can proceed via more complex pathways involving the nucleation of reversed domains and propagation of domain walls. These complex pathways are expected to introduce entropic contributions \cite{desplat2020entropy}, which can significantly influence relaxation times and attempt frequencies, presenting an interesting avenue for future investigation.

The strong strain sensitivity of the energy barriers in Mn$_3$Sn elucidated in this work suggests the possibility of voltage control over magnetic fluctuations. This can be achieved by integrating the noncollinear antiferromagnet with a piezoelectric substrate, forming a magnet/piezoelectric hybrid \cite{guo2020giant, chen2019electric, cai2017electric} structure that allows for dynamic electrical modulation of these fluctuations.

The capability to tune fluctuation rates over orders of magnitude using modest, experimentally achievable strains allows for the on-demand generation of random telegraph noise with tunable cutoff frequencies. Networks of such probabilistic bits (p-bits), potentially designed with a distribution of strain-controlled fluctuation rates, could be engineered to produce $1/f$-like noise spectra \cite{costanzi2017emergent}, a feature suggested as a resource for accelerating certain optimization algorithms \cite{du2022synaptic, eberhard2023pink}. 

Beyond stochastic computing, the strong magneto-elastic coupling evident from the strain-sensitive exchange and anisotropy parameters also points towards novel applications in antiferromagnetic magnonics \cite{chumak2015magnon}. While the direct electrical generation of magnons typically involves mechanisms such as spin-transfer torque or power-hungry dipolar fields \cite{liu2025correlated, barman20212021, chumak2015magnon}, dynamic strain---for instance, via surface acoustic waves on a piezoelectric substrate or time-varying gate voltages inducing local strain---can potentially coherently modulate magnon dispersion, and their interactions \cite{chai2024voltage, rongione2023emission, babu2020interaction, liao2024hybrid, xu2020nonreciprocal, wei2024strain, xu2020nonreciprocal}. Furthermore, patterning periodic arrays of electrodes on a piezoelectric substrate to create spatially modulated strain fields in an overlying Mn$_3$Sn film could realize an electrically tunable magnonic crystal. Such structures would allow for the engineering of magnon band structures, potentially leading to AFM-based THz filters, waveguides, and other advanced signal processing elements, thereby leveraging the high frequencies and inherent robustness of antiferromagnetic magnons.

\section{Conclusion}
In this work we systematically investigate the influence of epitaxial strain on the magnetic properties and fluctuation dynamics of Mn$_3$Sn noncollinear antiferromagnets. Through detailed DFT calculations, we have quantified the strain dependence of crucial magnetic parameters, revealing a significant impact on magnetic anisotropy, bilinear exchange, and particularly the biquadratic exchange interaction.  Our analysis shows that the interplay between these strain-modified interactions, especially the competition involving anisotropic Heisenberg and biquadratic exchange terms, dictates the thermal stability of these materials.  The derived effective model for octupole dynamics successfully connects these microscopic changes to the macroscopic fluctuation behavior, offering explanations for experimentally observed strain-induced anisotropies and thermal stability factors.  Beyond understanding current experimental results, our findings open avenues for actively controlling magnetic fluctuations via piezoelectric strain, suggesting pathways for realizing dynamically reconfigurable probabilistic circuits and novel magnonic devices. Future work could extend this framework to explore more complex reversal mechanisms and the role of entropic contributions in larger systems.

\section{Data availability}
All data will be made available by the authors upon reasonable request.
\appendix
\section{Appendix I: Details of DFT calculations}
We perform first-principles calculations based on density functional theory (DFT) using the \textsc{Quantum ESPRESSO} package~\cite{giannozzi2009quantum, giannozzi2017advanced}. The Perdew-Burke-Ernzerhof exchange-correlation functional within the generalized gradient approximation is employed~\cite{wu2006more}. A plane-wave basis set with energy cutoffs of 952.40~eV (70.00~$Ry$) and 10612.45~eV (780.00~$Ry$) for wavefunctions and charge density, respectively, is used. A $\Gamma$-centered $16 \times 16 \times 16$ $k$-point grid is used, and self-consistent calculations are performed until the total energy change between successive steps is less than $1.36 \times 10^{-5}$~eV ($10^{-6}$~$Ry$), ensuring convergence. Lattice vectors and atomic positions are optimized until forces are below 0.257~eV/nm (0.001~$Ry$/$a_0$), where $a_0= 0.0529$~nm. Noncollinear spin calculations with constrained atomic magnetization directions are used to compute energies for various magnetic states, generating the dataset for the energy mapping scheme.
\section{Appendix II: Fitting exchange parameters from DFT}
\begin{figure}[ht]
  \centering
  \includegraphics[width=0.45\textwidth]{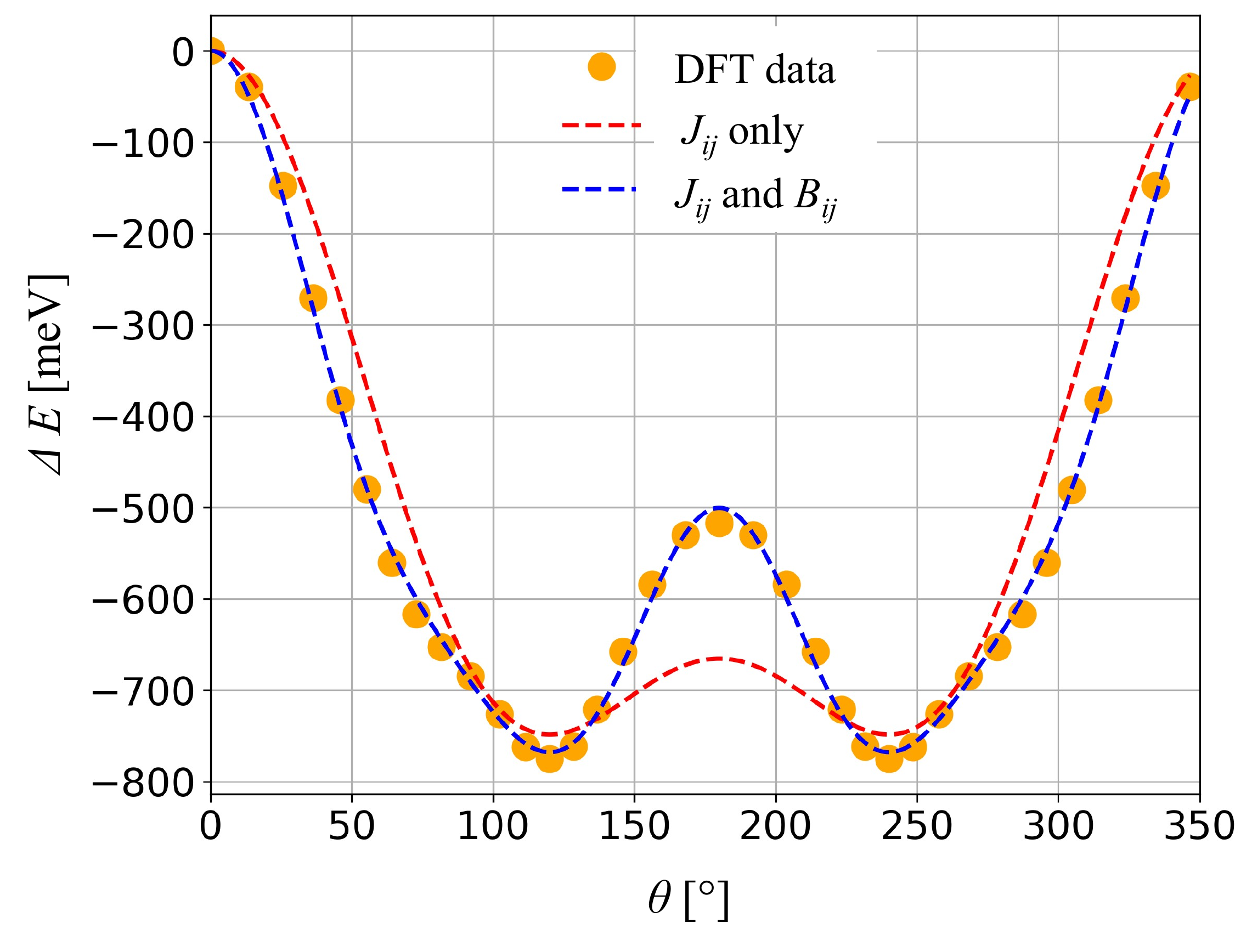}
\caption{Importance of biquadratic exchange in modeling the magnetic energy landscape of Mn$_3$Sn. The plot shows the change in DFT-calculated total energy ($\Delta E = E - E_0$) as a function of the relative orientation angle ($\theta$) of selected Mn spins (e.g., single-spin rotation from Fig.~\ref{fig:DFT_mapping_schematic}a). A fit using only bilinear Heisenberg exchange (red dashed line) fails to reproduce the DFT data (orange points), while a model including both bilinear and biquadratic exchange terms (blue dashed line) shows excellent agreement, underscoring the necessity of the biquadratic term. $E_0$ is the energy of the reference spin configuration which has been subtracted for all configurations.}
\label{fig:DFT_exchange_fit}
\end{figure}
As shown in Fig.~\ref{fig:DFT_exchange_fit}, the evolution of the total energy obtained from DFT cannot be captured by a Heisenberg term alone, which fails to reproduce the pronounced curvature of $\Delta E(\theta)$. By introducing a biquadratic exchange term, the fitted Hamiltonian achieves near‐perfect agreement with the DFT data over the full range of spin orientations. The reference energy $E_{0}$ was chosen as the energy of the initial spin configuration and subtracted from all points to highlight the relative changes. On the basis of this analysis, the biquadratic exchange contribution has been included in all subsequent calculations in this work. The results obtained from fitting for the unstrained case and a representative strained case (-1~\%) are summarized in the following tables. 
\begin{table}[H]
\caption{Composite bilinear and biquadratic exchange between Mn pairs (unstrained) for $q=0$ excitations. Each interaction between two atomic sites is the sum of interactions between one atomic site in one unit cell and the other atomic site and its equivalents in all unit cells. The distance indicated by $1^{\mathrm{st}}$ nearest-neighbor (NN) is the closest among all pairs, and $2^{\mathrm{nd}}$ NN is the next closest.}
\centering
\begin{tabular}{|c|c|c|c|c|}
\hline
\textbf{Pair} & $J_{ij}$ & $B_{ij}$ & $1^{\mathrm{st}}$ NN & $2^{\mathrm{nd}}$ NN \\
              & (meV)    & (meV)    & (nm)               & (nm)               \\ \hline
(2,3) & $16.82\pm0.07$  & $31.41\pm0.14$   & 0.2716 & 0.2875 \\ \hline
(5,6) & $16.82\pm0.07$  & $31.41\pm0.14$   & 0.2716 & 0.2875 \\ \hline
(1,2) & $16.82\pm0.07$  & $31.41\pm0.14$   & 0.2716 & 0.2875 \\ \hline
(1,3) & $16.82\pm0.07$  & $31.41\pm0.14$   & 0.2716 & 0.2875 \\ \hline
(4,5) & $16.82\pm0.07$  & $31.41\pm0.14$   & 0.2716 & 0.2875 \\ \hline
(4,6) & $16.82\pm0.07$  & $31.41\pm0.14$   & 0.2716 & 0.2875 \\ \hline
(1,5) & $45.42\pm0.08$  & $16.20\pm0.16$   & 0.2744 & 0.4858 \\ \hline
(1,6) & $45.42\pm0.08$  & $16.20\pm0.16$   & 0.2744 & 0.4858 \\ \hline
(2,4) & $45.42\pm0.08$  & $16.20\pm0.16$   & 0.2744 & 0.4858 \\ \hline
(3,4) & $45.42\pm0.08$  & $16.20\pm0.16$   & 0.2744 & 0.4858 \\ \hline
(2,6) & $45.42\pm0.08$  & $16.20\pm0.16$   & 0.2744 & 0.4858 \\ \hline
(3,5) & $45.42\pm0.08$  & $16.20\pm0.16$   & 0.2744 & 0.4858 \\ \hline
(2,5) & $-67.03\pm0.07$ & $-3.76\pm0.14$   & 0.3861 & 0.3973 \\ \hline
(3,6) & $-67.03\pm0.07$ & $-3.76\pm0.14$   & 0.3861 & 0.3973 \\ \hline
(1,4) & $-67.03\pm0.07$ & $-3.76\pm0.14$   & 0.3861 & 0.3973 \\ \hline
\end{tabular}
\label{tab:Jij_Bij_PairDistances}
\end{table}
\begin{table}[H]
\caption{Composite bilinear and biquadratic exchange between Mn pairs (1~\% strain) for $q=0$ excitations. Each interaction between two atomic sites is the sum of interactions between one atomic site in one unit cell and the other atomic site and its equivalents in all unit cells. The distance indicated by $1^{\mathrm{st}}$ nearest-neighbor (NN) is the closest among all pairs, and $2^{\mathrm{nd}}$ NN is the next closest.}
\centering
\begin{tabular}{|c|c|c|c|c|}
\hline
\textbf{Pair} & $J_{ij}$ & $B_{ij}$ & $1^{\mathrm{st}}$ NN & $2^{\mathrm{nd}}$ NN \\
              & (meV)    & (meV)    & (nm)               & (nm)               \\ \hline
(2,3) & $19.89\pm0.13$  & $33.09\pm0.25$  & 0.2689 & 0.2846 \\ \hline
(5,6) & $19.89\pm0.13$  & $33.09\pm0.25$  & 0.2689 & 0.2846 \\ \hline
(1,2) & $16.87\pm0.09$  & $30.19\pm0.17$  & 0.2709 & 0.2868 \\ \hline
(1,3) & $16.87\pm0.09$  & $30.19\pm0.17$  & 0.2709 & 0.2868 \\ \hline
(4,5) & $16.87\pm0.09$  & $30.19\pm0.17$  & 0.2709 & 0.2868 \\ \hline
(4,6) & $16.87\pm0.09$  & $30.19\pm0.17$  & 0.2709 & 0.2868 \\ \hline
(1,5) & $43.86\pm0.09$  & $17.35\pm0.18$  & 0.2756 & 0.4832 \\ \hline
(1,6) & $43.86\pm0.09$  & $17.35\pm0.18$  & 0.2756 & 0.4832 \\ \hline
(2,4) & $43.86\pm0.09$  & $17.35\pm0.18$  & 0.2756 & 0.4832 \\ \hline
(3,4) & $43.86\pm0.09$  & $17.35\pm0.18$  & 0.2756 & 0.4832 \\ \hline
(2,6) & $44.79\pm0.12$  & $17.30\pm0.25$  & 0.2762 & 0.4852 \\ \hline
(3,5) & $44.79\pm0.12$  & $17.30\pm0.25$  & 0.2762 & 0.4852 \\ \hline
(2,5) & $-67.02\pm0.09$ & $-3.67\pm0.18$  & 0.3966 & 0.3855 \\ \hline
(3,6) & $-67.02\pm0.09$ & $-3.67\pm0.18$  & 0.3966 & 0.3855 \\ \hline
(1,4) & $-66.57\pm0.13$ & $-1.95\pm0.27$  & 0.3967 & 0.3874 \\ \hline
\end{tabular}
\label{tab:Jij_Bij_strain_NN}
\end{table}
The uncertainties here indicate single standard deviations computed from the fits to the energies computed as a function of atomic moment orientations. 
%
\section{Appendix III: Perturbative theory of octupole dynamics}
\begin{figure}[ht]
\includegraphics[width=0.3\textwidth]{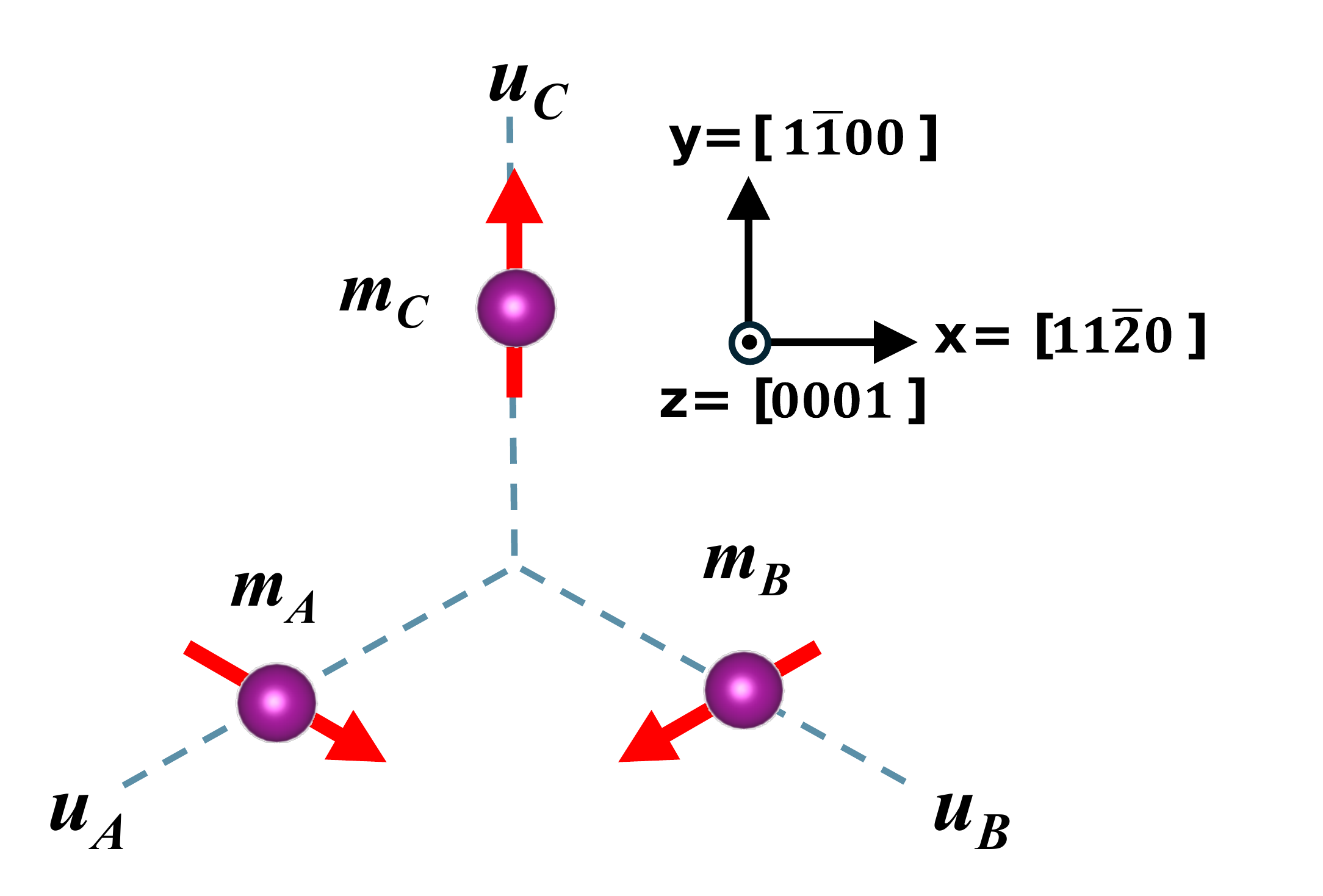}
\caption{Schematic representation of the effective three-spin system (macrospins A, B, C) derived from the original six-atom sublattice. The coordinate system and relevant anisotropy axes $u_A$, $u_B$ and $u_C$ used in the Hamiltonian are indicated. This simplified model captures the essential low-energy dynamics relevant to this study.}
\label{fig:theory_coord}
\end{figure}
This section details the derivation of the effective equations of motion that govern the collective dynamics of the cluster octupole moment in Mn$_3$Sn. For this purpose, we adapt an approach similar to those employed in Refs~\cite{he2024magnetic, konakanchi2025electrically} extended to account for the bilinear exchange terms. As established in the main text, the low-energy physics and the slower dynamical modes relevant to octupolar switching are often dominated by collective excitations. We leverage this by employing a reduced three-spin model. This simplification is justified by the presence of strong effective ferromagnetic coupling between specific pairs of Mn spins across different Kagome layers—a feature consistent with our DFT-extracted inter-layer exchange parameters (e.g., $J_{1,4} \approx -67.03$~meV, see Table~\ref{tab:Jij_Bij_PairDistances}) and also adopted in other literature~\cite{liu2017anomalous, he2024magnetic}. These pairings (specifically, atoms 1 with 4, 2 with 5, and 3 with 6, as per Fig.~\ref{fig:structure}(b)) lead to the formation of three effective `macrospins' (denoted A, B, and C; see Fig.~\ref{fig:theory_coord} for a schematic representation) each consisting of 6$\mu_B$ magnetic moment. This coarse-graining procedure effectively integrates out the faster, higher-energy modes that would arise from the independent dynamics of the six individual Mn spins, allowing us to focus on the collective dynamics pertinent to this work. The Hamiltonian density, $\mathcal{H}$, for this effective three-macrospin system, computed by converting the atomistic Hamiltonian to its corresponding micromagnetic quantities, can be expressed as
\begin{align}
\mathcal{H} =\,& J\bigl(\bm{m}_A\cdot\bm{m}_B + \bm{m}_B\cdot\bm{m}_C + \bm{m}_C\cdot\bm{m}_A\bigr) \nonumber \\
&+ D\,\bm{u}_z \cdot \bigl(\bm{m}_A\times\bm{m}_B + \bm{m}_B\times\bm{m}_C + \bm{m}_C\times\bm{m}_A\bigr) \nonumber \\
&- K\Bigl[ (\bm{m}_A\cdot\bm{u}_{aA})^2 + (\bm{m}_B\cdot\bm{u}_{aB})^2 + (\bm{m}_C\cdot\bm{u}_{aC})^2 \Bigr] \nonumber \\
&+ B\Bigl[ (\bm{m}_A\cdot\bm{m}_B)^2 + (\bm{m}_B\cdot\bm{m}_C)^2 + (\bm{m}_C\cdot\bm{m}_A)^2 \Bigr] \nonumber \\
&- \Delta B (\bm{m}_A\cdot\bm{m}_B)^2 - \Delta J  (\bm{m}_A\cdot\bm{m}_B)
\end{align}
Here, $\bm{m}_i$ (for $i=A, B, C$) represents the normalized magnetic moment vector of the $i$-th macrospin, whose orientation can be parameterized in spherical coordinates $(\theta_i, \phi_i)$ as $\bm{m}_i = (\cos\phi_i\sin\theta_i, \sin\phi_i\sin\theta_i, \cos\theta_i)$. The Hamiltonian incorporates several key interactions: isotropic Heisenberg exchange ($J$) and biquadratic exchange ($B$), Dzyaloshinskii-Moriya interaction (DMI, $D$) along $\bm{u}_z$, site-dependent uniaxial magnetocrystalline anisotropy ($K$) favoring alignment along local easy axes $\bm{u}_{ai}$, and strain-induced symmetry breaking captured by anisotropic exchange terms proportional to $\Delta J$ and $\Delta B$. 

The parameters in this effective model are derived from the microscopic interactions in Eq.~\eqref{eq:F0Hamiltonian}. The strain direction breaks the symmetry such that the interaction between macrospins A and B differs from the B-C and C-A interactions. The effective exchange parameters are defined by summing the relevant microscopic interactions:
\begin{align*}
    J_{BC} &= J = J_{23} + J_{26} + J_{35} + J_{56} \\
    J_{CA} &= J = J_{13} + J_{34} + J_{16} + J_{46}\\
    J_{AB} &= J - \Delta J =  J_{12} + J_{15} + J_{24} + J_{45} \\
\end{align*}
where the atom indices on the right correspond to those in Fig.~\ref{fig:structure}(b). An analogous set of transformations applies to the biquadratic exchange parameters $B$. 

Because each macrospin in the three‐spin model represents a pair of original Mn moments, all single‐site couplings inherited from the six‐spin Hamiltonian are doubled.  In particular, the uniaxial anisotropy constant becomes  
\begin{align}
  K = 2 K_1, 
\end{align}
and the isotropic DMI parameter entering Eq.~\eqref{eq:F0Hamiltonian} is  
\begin{align}
  D= 2 D_{\rm atomistic},
\end{align}
where $D_{\rm atomistic}$ corresponds to an intralayer coupling primarily responsible for choosing the chirality of the triangular spin order $(D_{12}=D_{23}=D_{31}=D/2)$ (and similarly on the other layer). For our numerical calculations, we used a value of $D_{atomistic}=12.7$ meV based on our preliminary DFT calculations. The site-dependent anisotropy axes $\bm{u}_{ai}$ for the macrospins lie within the $xy$-plane are found to be along
\begin{align} \label{eq:anisotropy_axes_app} 
    \bm{u}_{A} &= \left(-\tfrac{\sqrt{3}}{2}, -\tfrac{1}{2},  0\right), \quad
    \bm{u}_{B} = \left(-\tfrac{\sqrt{3}}{2},\tfrac{1}{2},  0\right), \quad 
    \bm{u}_{C} = (0, 1, 0)
\end{align}
To effectively analyze the dynamics, particularly to distinguish between the slow, collective (acoustic) modes of the octupole and the faster, internal (optical) modes of the macrospin structure, we transform from individual macrospin coordinates to a set of collective coordinates. These include a uniform azimuthal angle $\phi_0$ and a uniform out-of-plane canting angle $\theta_0$ for the entire triangular structure, along with relative angles ($\delta_A, \delta_B$) and ($u_A, u_B$) that describe deviations from this uniform state.
\begin{align} \label{eq:phi_transform_app} 
    \phi_0 &= -\frac{\phi_A + \phi_B + \phi_C}{3} \\
    \delta_A &= \frac{2\pi}{3} + \phi_A + \phi_0 \\
    \delta_B &= -\frac{2\pi}{3} + \phi_B + \phi_0
\end{align}
and the polar transformations
\begin{align} \label{eq:theta_transform_app} 
    \theta_0 &= \frac{\theta_A + \theta_B + \theta_C}{3} - \frac{\pi}{2} \\
    u_A &= \theta_A - \theta_0 - \frac{\pi}{2} \\
    u_B &= \theta_B - \theta_0 - \frac{\pi}{2}
\end{align}
In this new coordinate system, $\phi_0$ and $\theta_0$ represent the slow (acoustic) modes corresponding to the collective reorientation and canting of the octupolar order, respectively \cite{he2024magnetic}. Conversely, $\delta_A, \delta_B, u_A,$ and $u_B$ describe the fast (optical) fluctuations around the near-planar equilibrium configuration of the macrospins. To make the problem analytically tractable, we expand the Hamiltonian $\mathcal{H}$ up to the second order in these small fluctuation variables ($\delta_A, \delta_B, u_A, u_B,$ and $\theta_0$). Furthermore, consistent with the hierarchy of energy scales in Mn$_3$Sn where exchange interactions dominate, the anisotropy $K$ and the strain-induced exchange anisotropies ($\Delta J, \Delta B$) are treated as perturbations. This expansion yields:
\begin{widetext}
\begin{align}
\mathcal{H} \approx \frac{1}{2} \Bigg(
& \left( \sqrt{3} (-\Delta B + \Delta J)
- \sqrt{3} K \cos\left(2 \phi_o\right)
+ 3 K \sin\left(2 \phi_o\right) \right) \delta_A
+ (6 B + 3 J) \delta_A^2 \nonumber \\ 
& + \sqrt{3} (\Delta B - \Delta J + K \cos\left(2 \phi_o\right)) \delta_B 
+ 3 K \sin\left(2 \phi_o\right) \delta_B \nonumber \\
& + (6 B + 3 J) \delta_A \delta_B
+ (6 B + 3 J) \delta_B^2 \nonumber \\
& + (9 J - 9 B + 3\sqrt{3}D_z) \theta_o^2 \nonumber \\ 
& + \sqrt{3} D_z \Big( 2 \left(u_A^2 + u_A u_B + u_B^2\right)
+ 3 \left(\delta_A^2 + \delta_A \delta_B + \delta_B^2\right) \Big)
\Bigg) \label{eq:H_approx_app} 
\end{align}
\end{widetext}

With the approximated Hamiltonian, our objective is to derive the equations of motion for all dynamical variables $q_i \in \{\phi_0, \theta_0, \delta_A, \delta_B, u_A, u_B\}$. We employ the Euler-Lagrange formalism, constructing the Lagrangian $\mathcal{L} = \mathcal{L}_B - \mathcal{H}$. Here, $\mathcal{L}_B$ is the crucial kinetic term arising from the spin Berry phase ~\cite{dasgupta2020theory}. For our system of coupled macrospins, $\mathcal{L}_B$ can be written as
\begin{align*}
\mathcal{L}_B = \frac{\mu_0 M_S}{\gamma}\Big( (2 u_A + u_B) \dot{\delta}_A
+ (u_A + 2 u_B) \dot{\delta}_B  - 3 \theta_o \dot{\phi}_o \Big)
\end{align*}
where $M_S$ is the saturation magnetization, $\gamma$ is the gyromagnetic ratio. To incorporate energy dissipation due to damping processes, we introduce the Rayleigh dissipation function $\mathcal{R}$
\begin{align}
\mathcal{R}  = \frac{\alpha \mu_0 M_S}{\gamma} \Big(\dot{u}_A^2 &+ \dot{u}_A \dot{u}_B + \dot{u}_B^2 + \dot{\delta}_A^2 + \dot{\delta}_A \dot{\delta}_B\\ &+ \dot{\delta}_B^2  + \tfrac{3}{2} \left( \dot{\theta}_o^2 + \dot{\phi}_o^2 \right)
\Big)
\end{align}
where $\alpha$ is the Gilbert damping parameter, phenomenologically accounting for the relaxation mechanisms. Applying the Euler-Lagrange equations yields the following set of coupled differential equations for the system's dynamics
\begin{widetext}
\begin{align}
\label{eq:eom_phi0_full} \sqrt{3}\,K\,\sin(2\phi_0)\Bigl(-\delta_A+\delta_B\Bigr)
-3\,K\,\cos(2\phi_0)\Bigl(\delta_A+\delta_B\Bigr)
+3\frac{\mu_0 M_S}{\gamma} \Bigl(\dot{\theta}_0-\alpha\,\dot{\phi}_0\Bigr)&=0, 
\end{align}
\begin{align}
\label{eq:eom_theta0_full} 
-\Bigl(9\,J - 9\,B -3\sqrt{3}\,D\Bigr)\theta_0
-3\alpha\,\frac{M_S}{\gamma} \,\dot{\theta_0}
-3\frac{\mu_0 M_S}{\gamma} \,\dot{\phi_0}&=0,
\end{align}
\begin{align}
\label{eq:eom_deltaA_full} 
\sqrt{3}\,(\Delta B - \Delta J)
+\sqrt{3}\,K\,&\cos(2\phi_0)
-3\,K\,\sin(2\phi_0)
-3\Bigl(2B+\sqrt{3}\,D+J\Bigr)
\Bigl(2\delta_A+\delta_B\Bigr) \notag\\
&-2\frac{\mu_0 M_S}{\gamma} \,\Bigl(2\dot{u}_A+\dot{u}_B+\alpha\bigl(2\dot{\delta}_A+\dot{\delta}_B\bigr)\Bigr)=0,
\end{align}
\begin{align}
\label{eq:eom_deltaB_full} 
\sqrt{3}\,(-\Delta B + \Delta J)
-\sqrt{3}\,K\,&\cos(2\phi_0)
-3\,K\,\sin(2\phi_0)
-3\Bigl(2B+\sqrt{3}\,D+J\Bigr)
\Bigl(\delta_A+2\delta_B\Bigr) \notag\\
&-2\frac{\mu_0 M_S}{\gamma} \,\Bigl(\dot{u}_A+2\dot{u}_B+\alpha\bigl(\dot{\delta}_A+2\dot{\delta}_B\bigr)\Bigr)=0,
\end{align}
\begin{align}
\label{eq:eom_uA_full} 
-\sqrt{3}\,D\Bigl(2u_A+u_B\Bigr)
+\frac{\mu_0 M_S}{\gamma} \,\Bigl[-\alpha\Bigl(2\dot{u}_A+\dot{u}_B\Bigr)+2\dot{\delta}_A+\dot{\delta}_B\Bigr]&=0,
\end{align}
\begin{align}
\label{eq:eom_uB_full} 
-\sqrt{3}\,D\Bigl(u_A+2u_B\Bigr)
+\frac{\mu_0 M_S}{\gamma} \,\Bigl[-\alpha\Bigl(\dot{u}_A+2\dot{u}_B\Bigr)+\dot{\delta}_A+2\dot{\delta}_B\Bigr]&=0.
\end{align}
\end{widetext}
This set of six coupled equations provides a complete description of the dynamics of all collective coordinates. As previously identified, $(\phi_0, \theta_0)$ represent the slow collective modes of the octupole, while $(\delta_A, \delta_B, u_A, u_B)$ describe the fast internal modes of the macrospin structure. Given the typical separation in timescales between these acoustic and optical modes, we can simplify the description by employing adiabatic elimination. This involves setting the time derivatives of the fast modes to zero ($\dot{\delta}_A, \dot{\delta}_B, \dot{u}_A, \dot{u}_B = 0$) in Eqs.~\eqref{eq:eom_deltaA_full}-\eqref{eq:eom_uB_full}, solving for these fast variables in terms of the slow variables, and substituting these expressions back into the equations for the slow modes (Eqs.~\eqref{eq:eom_phi0_full}-\eqref{eq:eom_theta0_full}). This procedure effectively integrates out the fast degrees of freedom, yielding a reduced set of equations that govern the low-energy dynamics of the octupolar order parameters $\phi_0(t)$ and $\theta_0(t)$. Finally, for small canting of spins out of the $xy$ plane, the out-of-plane projection of the dipole moments formed by macrospins A, B and C is $m_z \approx -\theta_0$. Consequently
\begin{equation} \label{eq:final_eom_phi0_app} 
\omega_K^{\text{eff}}\sin(2\phi_0)
-\dot{m}_z
-\alpha \dot{\phi}_0=0,
\end{equation}
\begin{equation} \label{eq:final_eom_theta0_app} 
\omega_{\perp}^{\text{eff}}m_z +\Bigl(\alpha\,\dot{m}_z-\dot{\phi}_0\Bigr)=0.
\end{equation}
These compact equations describe the essential low-energy dynamics of the octupole moment. The effective parameters $\omega_K^{\text{eff}}$ and $\omega_{\perp}$ are defined as
\begin{align}
    \omega_K^{\text{eff}} &= \frac{2\gamma K(-\Delta B + \Delta J)}{3\mu_0 M_S (2B + \sqrt{3}D + J)} \label{eq:omega_K_eff_app} \\
    \omega_{\perp}^{\text{eff}} &= \frac{\gamma (3J - 3B + \sqrt{3}D)}{\mu_0 M_S} \label{eq:omega_perp_app}
\end{align}

\section{Appendix IV: Numerical Solution of the Stochastic Landau-Lifshitz-Gilbert Equation}
To benchmark our analytical models, we performed numerical simulations using the stochastic Landau-Lifshitz-Gilbert equation.
The equation of motion for the normalized magnetic moment $\bm{m}_i = \bm{M}_i/M_S$ of spin $i$ is given by
\begin{equation}
    \label{eq:LLGeqn_appendix}
    \frac{d\bm{m}_i}{dt} = -\gamma \bm{m}_i \times (\bm{H}_{\text{eff},i} + \bm{H}_{\text{th},i}) + \alpha \bm{m}_i \times \frac{d\bm{m}_i}{dt}
\end{equation}
where $\gamma$ is the gyromagnetic ratio, and $\alpha$ is the Gilbert damping parameter. The effective field $\bm{H}_{\text{eff},i}$ incorporates the exchange, magnetocrystalline anisotropy,  and DMI. $\bm{H}_{\text{th},i}(t)$ is the stochastic thermal field obeying the following statistical properties 
\begin{align}
    \langle H_{\text{th},i}^{\mu}(t) \rangle &= 0 \\
    \langle H_{\text{th},i}^{\mu}(t) H_{\text{th},j}^{\nu}(t') \rangle &= \frac{2 \alpha k T}{\gamma \mu_0 M_S V} \delta_{ij} \delta_{\mu\nu} \delta(t-t')
\end{align}
where $V$ is the magnetic volume of the spin. The simulations were carried out in the six spin space. The dynamics of each spin's orientation $(\theta_i(t), \phi_i(t))$ were integrated using an in-house Python code employing a second-order Heun's solver. Unless otherwise noted, all sLLG simulations were performed at temperature $T = 300$~K with Gilbert damping $\alpha = 0.05$. A time step of $dt = 0.1 \text{ ps}$ was chosen based on convergence tests to ensure numerical stability and accuracy. Trajectories of $(\theta_i(t), \phi_i(t))$ for all Mn spins were recorded. These numerical results were compared against analytical expressions from Langer's theory and the zero-barrier limit approximation.

\section{Appendix V: Magnon Spectrum}
\label{sec:magnon}
\begin{figure}[htbp]
  \centering
  \includegraphics[width=0.99\columnwidth]{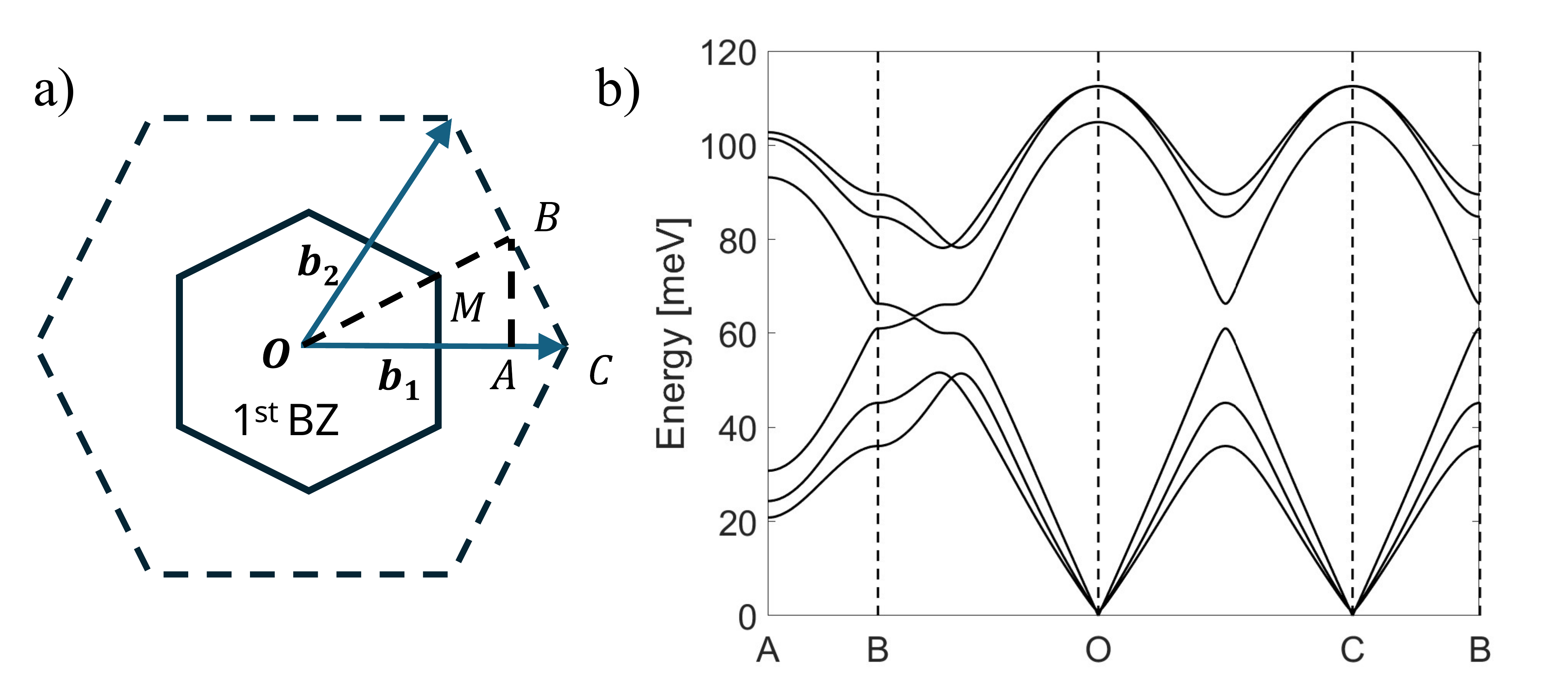} 
  \caption{(a) High-symmetry path A $\rightarrow$ B $\rightarrow$ $\Gamma$ $\rightarrow$ C $\rightarrow$ B in the Brillouin zone used for the magnon dispersion calculation. (b) Calculated magnon spectrum of unstrained Mn$_3$Sn along the specified high-symmetry path using the DFT-derived Hamiltonian and linear spin wave theory via the SpinW package. The spectrum qualitatively captures features observed in experimental studies (e.g., Park et al.~\cite{park2018magnetic}), though quantitative differences exist, potentially due to factors like the precise DMI parameters, exchange interactions beyond the $2 \times 1$ supercell considered for parameter extraction, or other effects not included in the current model.}
  \label{fig:magnon_spectrum}
\end{figure}

In this section, we utilize the first-principles DFT-extracted Hamiltonian parameters for unstrained Mn$_3$Sn to predict its magnon spectrum. To calculate the full magnon dispersion, it is necessary to resolve the composite exchange couplings, which are summed over all unit cells for the uniform mode analysis, into individual pairwise interactions. To achieve this, we performed additional DFT energy mapping calculations on 2x1x1 and 1x2x1 supercells of the primitive unstrained lattice. In particular, to map out the intra- and intercell contributions of the exchange, spins corresponding to specific bonds, including the first and second nearest-neighbor pairs of (1, 2) and (2, 5) (and their symmetry equivalents), were rotated independently. The corresponding changes in the self-consistent field energy were then fitted to the Hamiltonian using the existing exchange values. This procedure allows us to isolate the distinct intercell and intracell exchange coupling strengths required for the spin-wave analysis. With these individual pairwise interactions determined, we calculated the magnon spectrum using linear spin wave theory as implemented in the SpinW package~\cite{Toth2015linear}. The diagonalization of the Hamiltonian in reciprocal space yields the magnon dispersion relations. In Fig.~\ref{fig:magnon_spectrum}(b) we show the computed magnon spectrum along a high-symmetry path (depicted in Fig.~\ref{fig:magnon_spectrum}(a) within the hexagonal Brillouin zone). The spectrum clearly shows distinct acoustic and optical magnon branches, characteristic of multi-sublattice magnetic systems.

Our theoretical dispersion captures some qualitative trends observed in experimental inelastic neutron scattering data for unstrained Mn$_3$Sn, such as those reported by Park et al.~\cite{park2018magnetic}. However, quantitative mismatches exist. These discrepancies can be attributed to several factors, including the precise DMI vectors not fully resolved in this study, exchange interactions extending beyond the $2 \times 1$ supercell used for fitting, or other many-body effects beyond linear spin wave theory.

\bibliography{main.bib}
\end{document}